\documentclass[]{elsart}
\usepackage{epsfig}
\usepackage{lineno}
\begin{document}

\begin{frontmatter}

\title{Anisotropy studies around the galactic centre at EeV energies 
with the Auger Observatory}


\author{\bf The Pierre Auger Collaboration:}
\author{J.~Abraham$^{6}$,} 
\author{M.~Aglietta$^{41}$,} 
\author{C.~Aguirre$^{8}$,} 
\author{D.~Allard$^{74}$,} 
\author{I.~Allekotte$^{1}$,} 
\author{P.~Allison$^{70}$,}
\author{C.~Alvarez$^{45}$,} 
\author{J.~Alvarez-Mu\~{n}iz$^{59}$,} 
\author{M.~Ambrosio$^{38}$,}
% ...

\newpage

\begin{abstract}
Data from the Pierre Auger Observatory are
analyzed to search for anisotropies near the direction of the Galactic Centre
 at EeV energies. The exposure of 
 the surface array in this part of the sky is already significantly 
larger than that of 
the fore-runner experiments.  Our 
results do not support previous findings of localized excesses 
in the AGASA and  SUGAR data. We  
set an upper bound on a point-like  flux of cosmic rays 
arriving from the Galactic Centre which excludes several scenarios predicting
sources of EeV neutrons from Sagittarius $A$. 
 Also the events detected simultaneously by the surface 
 and fluorescence detectors (the `hybrid' data set), which
 have better pointing 
accuracy but are less numerous
 than those of the
surface array alone,  do not show  any significant localized
excess from this direction.
\end{abstract}

\end{frontmatter}

\section{Introduction}

The Galactic Centre region constitutes an attractive target for 
cosmic ray (CR) anisotropy studies at EeV energies, where 1~${\rm
  EeV}=10^{18}$~eV. These may be the highest energies for which the
galactic component of the cosmic rays is still dominant. 
Moreover, since the Galactic Centre (GC) 
harbors the very massive black hole associated with the radio source Sagittarius
$A^*$, as well as the expanding supernova remnant Sagittarius A East,  
it contains objects that might be candidates for powerful CR accelerators. 
The recent high significance observation by H.E.S.S. 
of a TeV $\gamma$ ray source near the 
location of  Sagittarius $A^*$ \cite{hess}, 
together with the discovery of a region 
of extended emission from
giant molecular clouds in the central 200~pc of the Milky Way \cite{hess2}, 
further motivates
 the search for excesses in this direction. 
The  location of the Pierre Auger Observatory in the southern hemisphere 
makes it particularly suitable
for anisotropy studies in this region since the GC, 
passing only $6^\circ$ from the
zenith at the site, lies well within the field of view of the
experiment. The number of CRs of EeV energies accumulated so
far at  the Pierre Auger Observatory 
from this part of the sky greatly
 exceeds that from previous observations, 
allowing  several interesting  searches to be made.

There have been reports by the 
AGASA experiment \cite{ha98,ha99} 
indicating a $4.5\sigma$ excess of cosmic rays 
 with energies in the range $10^{18}$--$10^{18.4}$~eV in a 
$20^\circ$ radius region centred at right ascension and declination
 coordinates  $(\alpha,\delta)\simeq
( 280^\circ,-17^\circ)$, 
in which the number of observed and expected events \cite{ha99} are $n_{obs}/n_{exp}
=506/413.6=1.22\pm 0.05$, where the error quoted is the one associated with
Poisson background fluctuations.
Note that the GC itself, for which we will adopt hereafter the 
Sagittarius~$A^*$ 
 J2000.0 coordinates, 
$(\alpha,\delta)=(266.3^\circ,-29.0^\circ)$, 
lies outside the AGASA field of view ($\delta>-24.2^\circ$).
Later searches near this region with a reanalysis of SUGAR data \cite{be01},
though with smaller statistics, failed to confirm these findings, 
but reported a  $2.9\sigma$ excess flux of CRs with energies in the range  
$10^{17.9}$--$10^{18.5}$~eV in a region of $5.5^\circ$ radius centred at 
$(\alpha,\delta)=(274^\circ,-22^\circ)$, for which they obtained 
$n_{obs}/n_{exp}=21.8/11.8=1.85\pm 0.29$. 

It is also sensible to search for a point-like excess from the GC.
Due to the imperfect reconstruction of the
arrival directions, the point
source would  be smeared 
on the angular scale of the resolution of the experiment.
In particular,  EeV 
neutrons emitted by one of the possible energetic sources
in the centre of the Galaxy may reach the Earth before decaying, 
and they would not be deflected by galactic magnetic fields.
 It is interesting to note that 
several scenarios predicting neutron fluxes from the GC detectable by Auger
have been put forward in recent years \cite{mt01,bo03,ah05,cr05,gr05,bi04}.

In this work we use Auger data from the on-going construction phase
to test the previous reports of localized excesses
 obtained with AGASA and SUGAR data,  and to set 
limits on a CR flux from the GC direction in a window matched to the angular  
resolution of the experiment at EeV energies. A preliminary analysis of this
kind was presented in \cite{le05}.

The AGASA experiment has also reported a large scale anisotropy at EeV
energies corresponding to a dipole-like modulation in
right ascension of $\sim 4$\% amplitude, 
with a maximum near the GC and a deficit in the anti-centre
direction. We defer the analysis of
such large scale signatures for future work. This will require, in particular, 
control of the systematic uncertainty of the
 modulation of the exposure in right ascension induced
by weather effects, 
which for the present Auger  data set  
is estimated to be at a level of 1\%. Uncertainties in
the background estimates at this level 
 do not affect the conclusions reached in the search for localized excesses
 performed in the present work.

\section{Data set}

The Auger surface detector \cite{ea04}, 
located in Malarg\"ue, Argentina (latitude
$-35.2^\circ$, longitude $69.5^\circ$ W  and mean altitude 1400 m a.s.l.), 
  has been growing in size during the data taking
period considered in this work, which goes
from January~$1^{st}$ 2004 (when 154 detectors had been deployed) 
 to  March~$30^{th}$ 2006 
(when 930 detectors were already deployed).
The surface detectors consist of plastic tanks filled with 12000 litres of
ultra-pure water in which the charged particles from the air showers 
produce Cherenkov light, which
is reflected by the Tyvek$^{\tt TM}$ liners and collected by three
photomultipliers.  The
basic cell of the array is triangular,  with separations of 1.5~km between
detector units, and hence 
the complete array with 1600 detectors will  cover an
area of 3000~km$^2$.

We consider the events from 
the surface detector (SD) array with three or more tanks
triggered in a compact configuration.  The events have
 to satisfy the level 5 quality trigger condition, which requires
 that the detector 
with the highest signal be surrounded by a hexagon of working detectors,
since this ensures that the
 event is well reconstructed. We also restrict the events
to zenith angles $\theta<60^\circ$.

The energies are obtained using the inferred signal size at 1000~m 
 from the
reconstructed shower core, $S(1000)$,
 adopting a conversion that leads to a constant flux
in different sky directions above 3~EeV, where the acceptance is saturated.
This is the so-called Constant Intensity Cut criterion implemented in
\cite{so05}. A calibration of the energies is performed 
using clean fluorescence data, i.e. hybrid events that were recorded when
 there were contemporaneous aerosol measurements, whose longitudinal profiles
 include the shower maximum in a measured range of at least 350~g~cm$^{-2}$ and in
 which there is less than 10\% Cherenkov
 contamination.  The estimated systematic 
uncertainty in the reconstructed shower
 energy with the fluorescence technique is currently 25\% \cite{be05}. 
For the hybrid events measured with both
techniques the dispersion between SD and FD energy assignments is  at
the level of 35\% in this energy range.
 From the uncertainty in the measurements of the signals from the
 Cherenkov tanks \cite{piera} 
the statistical uncertainty in the energy  determination which results from the
fitting procedure  is about 20\% for
the energy range considered  in this work, i.e. $10^{17.9}\
{\rm eV}<E<10^{18.5}$~eV.
 Notice that in this energy range
 48\% of the events involve just three tanks, 34\%
involve 4 tanks and only 18\%  more than 4  tanks. For three tank events the 
 68\% quantile  angular resolution is about  
$2.2^\circ$ and the resolution improves for events with 4 tanks or more
\cite{angres}.

Regarding the hybrid events, i.e. those with signal from 
 both the  fluorescence detectors 
(FD) and  surface array,  the 
angular resolution achieved is much smaller, typically below 1 degree
 \cite{angres}.  
Also, given that hybrid events may trigger with just one surface detector, 
the associated energy threshold ($\sim
10^{17}$~eV) is lower, and events up to zenith angles of
 $75^\circ$ are included in the data set.  However, the statistics accumulated
are significantly less, in part due to the $\sim 15$\% duty cycle of the
 fluorescence telescopes and
also because at EeV energies the FD is not fully efficient at
detecting showers over the full SD array.
There are for instance 79265 SD events in the data set considered
with energies  $10^{17.9}\ 
{\rm eV}<E<10^{18.5}$~eV, 
while the corresponding number of well reconstructed hybrid events 
in the same energy range is just 3439.
Note that $\sim 25$\% of the hybrid events in this energy range
involve less than three surface detectors, and are hence not included
in the SD only data set. 
\section{Results}

To study the possible presence of anisotropies, one needs first to
obtain the background expectations for the different sky directions under the
assumption of an isotropic CR distribution. 
This is a delicate issue since modulations of the exposure in right
ascension  are induced by 
the dead time of the detectors and the constantly growing array size. Also the
effects of weather variations, especially near the energy threshold of the 
detector, may be
 non-negligible since they may affect the shower development in the
atmosphere and/or the response of the  electronics. 
Preliminary studies of these effects indicate that the
possible weather-induced background modulations for the present data set
are at a level of 1\%, and are 
hence below the Poisson noise for the angular windows
considered\footnote{A detailed account of weather effects is certainly 
necessary to test
  large scale patterns at the few percent level. Relevant studies are in
  progress. }.

We have followed two different approaches \cite{ha05} 
to estimate the isotropic expectations for the SD analysis:
\begin{itemize}
\item {\bf The semi-analytic technique:}  At  EeV energies the zenith angle
  dependence of the exposure differs from the geometric one 
 corresponding to full acceptance, d$N\propto \sin\theta
  \cos\theta\ {\rm d}\theta$, mainly due to
  the attenuation in the atmosphere which 
affects large zenith angle showers. We therefore
  perform an analytic fit to the $\theta$ distribution of the observed events
  in the energy range under study and then make a convolution with the number
  of  hexagons with active detectors 
(which gives a measure of the aperture
   for events 
   satisfying the quality trigger criterion) as a function of time,
  assuming a uniform response in azimuth. Through this procedure one obtains
  an exposure which accounts
  for the non-saturated acceptance effects and for the non-uniform
  running times and array growth. 
 This technique allows to recover the detector's acceptance with negligible
biases even in the case in which a large scale pattern is present in the CRs
arrival distribution (see ref.~\cite{ha05} for details).

\item {\bf The shuffling technique:} Here the expected number of events in any
  direction is obtained by averaging many data sets obtained by shuffling the
  observed events in the energy range of interest so that the arrival times
  are exchanged among them and the azimuths are drawn uniformly. The
  shuffling can be performed in separate zenith angle bins or  by  just
mixing them
  all, and we found no significant difference between these two
  possibilities. By construction, 
this exposure preserves exactly the $\theta$ distribution of
  the events and accounts for the detector dead times, array growth and even
  in principle  for weather-induced modulations. It might however partially
  absorb modulations induced by large scale intrinsic anisotropies present 
in the CR
  flux, such as those due to a global dipole. 
\end{itemize}

 As implemented in the current
  analysis, the  two techniques differ
essentially in the treatment of
the time dependence of the detectors acceptance.
 With shuffling we follow the detected rates while with the semi-analytic
technique we assume a dependence only
proportional to the detector size, and these two quantities differ only
slightly.

The background estimate obtained with the shuffling technique in the GC region
turns out to be about 0.5\% larger than the one obtained with the semi-analytic
method. Since this difference is much smaller than the size of the
excesses that we are testing and is also below the level of the Poisson
fluctuations, we will hence  mainly quote
in the following the values obtained using the semi-analytic technique.

\subsection{Testing the AGASA and SUGAR excesses}

\begin{figure}[ht]
\centerline{{\epsfig{width=5.in,file=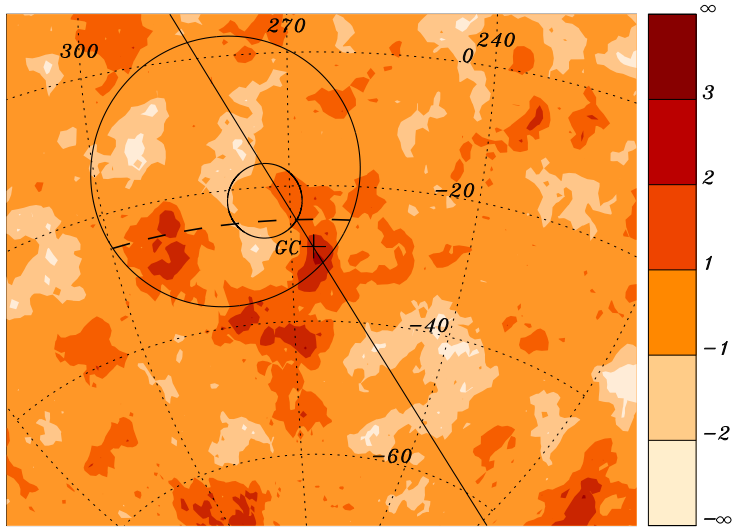}}}
\caption{Map of CR overdensity significances near the GC region on top-hat
  windows of $5^\circ$  radius. The GC location 
is indicated with a cross, lying along the galactic plane (solid line). 
Also the regions where the AGASA  experiment found their largest 
excess as well as the region of the SUGAR excess are indicated.}
\label{fitdip}
\end{figure}

In Figure~\ref{fitdip} we show a map of the GC region depicting  the 
Li-Ma significances\footnote{For the $\alpha$ parameter in the expression
  of the Li-Ma significance we use $\alpha=n_{exp}/n_{t}$, with
  $n_{t}$ the total number of events in the energy range considered and
  $n_{exp}$ the background expected in the angular region searched.} 
\cite{li83}   of overdensities 
in circular windows of $5^\circ$ degree radius, for SD data
with energies in the range $10^{17.9}$--$10^{18.5}$~eV. 
 This angular scale is convenient to visualize the distribution of
overdensities in
the windows explored by SUGAR and AGASA.
The galactic plane is represented
with a solid line and the location of the Galactic Centre is indicated with a
cross. The region in which AGASA reported an excess (in a slightly narrower
energy range)
is the big circle in the neighborhood of the GC, with the
dashed line indicating the lower boundary of the region observed by AGASA.
The smaller circle indicates the region where an excess in the SUGAR data was
reported.

\begin{figure}[ht]
\centerline{{\epsfig{width=3.in,angle=-90,file=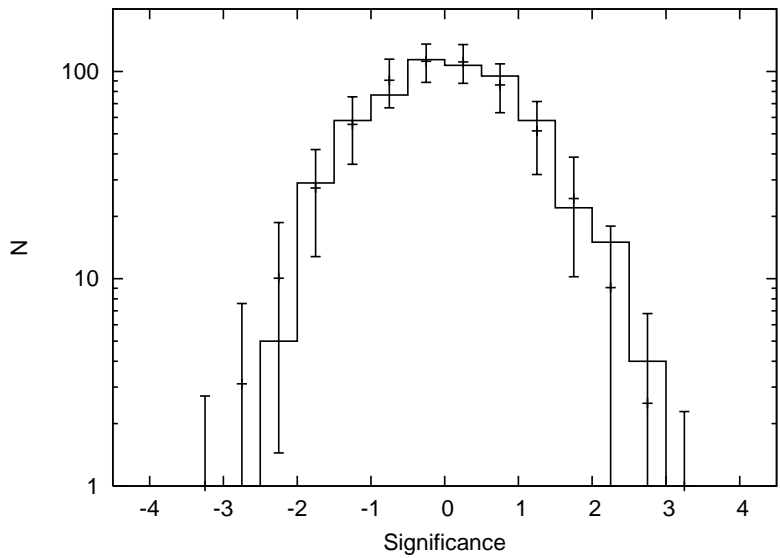}}}
\caption{Histogram  of overdensities on $5^\circ$
radius windows and for  $10^{17.9}\ 
{\rm eV}<E<10^{18.5}$~eV, together with isotropic expectations 
(average and $2\sigma$ bounds). Overdensities are computed on a grid of
$3^\circ$ spacing for the patch of the sky depicted in Fig.~\ref{fitdip}.}
\label{ovhisto}
\end{figure}

The size of the overdensities present in this map
is  consistent with what would be expected
as a result of statistical fluctuations of an isotropic sky. Indeed,
Figure~\ref{ovhisto} depicts the distribution of these
 overdensities  together with the expectations from an
isotropic flux (average and $2\sigma$ bounds obtained from Monte Carlo
simulations), and no significant departure from isotropy is observed.

For the $20^\circ$ circle centred at the AGASA location and for
$10^{18}\ {\rm eV}<E<10^{18.4}$~eV, 2116 events are
observed while 2159.6 are expected using the semi-analytic technique, 
while 2169.7
 are expected using the shuffling technique. It is clear that no
significant excess is observed. 
Note that the number of events is more than four times that collected by AGASA 
in this region,
in part due to the fact
that the GC lies well within the field of view of Auger, and in part due to
the fact that the total exposure of Auger is already double that achieved by
 AGASA.

\begin{table}[hb]
\centerline{\begin{tabular}{|c|c|c|}
\hline
${\rm E_{min}}$ [eV] &  ${\rm E_{max}}$ [eV]  &  $n_{obs}/n_{exp}$   \\
\hline
&& \\[-10pt]
$10^{17.9}$ & $10^{18.3}$  & $3179/3153.5= 1.01\pm 0.02(stat) \pm 0.01(syst)$ \\
\hline
&& \\[-10pt]
$10^{18}$ & $10^{18.4}$ &  $2116/2159.5 =0.98\pm 0.02(stat) \pm 0.01(syst)$ \\
\hline
&& \\[-10pt]
$10^{18.1}$ & $10^{18.5}$ &$1375/1394.5 = 0.99 \pm 0.03(stat) \pm 0.01(syst)$ \\
\hline
\end{tabular}}
\caption{Events in the AGASA region for different shifted energy intervals.}
\label{eshift.tab}
\end{table}

It must be borne in mind that there may be systematic differences in the
energy calibration of the two experiments.
To test whether these differences could have possibly masked the AGASA reported
excess, we show in Table~1 the observed and expected rates for different
energy ranges, offset by 0.1 decade in energy (i.e. by about 25\%),
keeping $E_{max}/E_{min}$ fixed. We have added a
systematic error of 1\% to the expected rates  to
account for the effects of possible weather induced modulations.
These results show that no significant excesses are seen in the AGASA region for
any of these cases. In particular, at the $2\sigma$ level the
excess in this region is always less than 6\%, well below the
 22\% excess reported by AGASA.

Since it is conceivable that  particles leading to a localized excess are
different from the bulk of the CRs (e.g. if they are nucleons and the bulk of
the CRs in this energy range are heavier nuclei), 
one may also wonder if the Auger sensitivity to these particles could be
reduced.  In particular, since for Auger 
the acceptance in this energy range is not yet saturated, 
it will be larger for heavy nuclei than for protons because
showers initiated by heavier primaries develop earlier and 
are hence more spread out at ground level. 
Using the estimates in \cite{all05} for the acceptance of p and Fe primaries,
we find that the sensitivity to protons is about $\sim 30$\% smaller 
than to Fe in the energy range studied (assuming an $E^{-3}$ spectrum).
In the case in which the 22\% 
excess reported by AGASA (which had full efficiency at EeV energies) was due to
nucleons while the background was due to heavy nuclei, at least a 15\%
excess should have been expected in Auger data. This  is 
much larger than the upper limit we are obtaining.

Regarding the localized excess observed in SUGAR data, we find in the 
same angular window and energy range that 
$n_{obs}/n_{exp}=286/289.7=0.98\pm 0.06$, 
and hence with 
more than an order of magnitude larger statistics no significant excess
is seen in this window. Shifting the energy range to account for
possible offsets also resulted in  no significant excess.

\subsection{Bounds on a point-like neutron source at the GC}

\subsubsection{The surface detector results}

The optimal  search for a point-like source is best done using a Gaussian 
filter matching the angular resolution of the experiment \cite{bi05}.
 For this 
we can assume that the reconstructed directions are distributed with respect
to the true direction (separated by an angle $\beta$) according to 
${\rm exp}(-\beta^2/2\sigma^2)$ per unit solid angle, 
where $\sigma\simeq 1.5^\circ$ at EeV energies, corresponding to a 
68\% quantile of $2.25^\circ$, where we have 
ignored a mild zenith angle dependence for simplicity.

We use for this search an energy range between $E_{min}=10^{17.9}$~eV
and $E_{max}=10^{18.5}$~eV. Below $E_{min}$
 the Auger SD acceptance is very suppressed. Note also 
that most neutrons from a source at the GC 
would have decayed in flight before reaching the Earth  for lower energies.
On the other hand, energies above $E_{max}$ may be hard to achieve for
galactic sources.

For the Gaussian window centred in the Sagittarius $A^*$ direction we get
$n_{obs}/n_{exp}= 53.8/45.8$. This corresponds to a ratio of $1.17\pm 0.10$,
where the estimate of the uncertainty takes into account that
the window is Gaussian.  Applying the results of 
\cite{bi05}, we
get a 95\% CL upper bound on the number of events from the source 
of $n_s^{95}=18.5$.
To translate this into a bound on the source flux we make two assumptions:
\begin{itemize}
\item We assume
 that the spectrum of the source is similar to that of the CRs,
which is approximately  $\propto E^{-3}$ in this energy range. 
If the source spectrum were
actually harder, the bound we obtain would be a conservative one.
\item
We assume that the composition of the CRs in this energy range
is similar to that of the source, i.e. proton-like. We will then discuss how
the limit is modified if the CRs were heavier, in which case the detector
acceptance would be different for the bulk of the 
CRs and for the neutron source.
\end{itemize}

Under these assumptions,  the energy dependent acceptance of the
detector has the same effect upon the source flux and the background flux, so
that one can  relate the ratio between the CR flux
and the expected number of background events in this window, with the ratio
between the source flux upper limit and the bound obtained for $n_s^{95}$. 

We take for the differential CR spectrum flux the expression
\begin{equation}
\Phi_{CR}(E)\simeq \kappa\; 50 \left(\frac{E}{\rm
  EeV}\right)^{-3.3} {\rm EeV^{-1}\ km^{-2}\ yr^{-1}\ sr^{-1}},
\end{equation}
which has an $E^{-3.3}$ dependence (consistent with the
value found e.g. by HiRes \cite{be05b} in the energy range $10^{17.5}
\ {\rm eV}<E<10^{18.5}$~eV),  and is a smooth
extrapolation of   the spectrum measured at the
Auger Observatory\footnote{A power law fit to the Auger Observatory
measurements
  \cite{so05}
leads to
$\Phi_{CR}(E)=(30.9\pm 1.7)\times (E/{\rm EeV})^{-2.84\pm 0.03}{\rm EeV^{-1}\
    km^{-2}\ yr^{-1}\ sr^{-1}}$  (statistical error only).} at $E>3$ EeV.

The factor $\kappa$ is introduced
 to parametrise our limited
knowledge of the true CR flux and it should be of order unity according to 
the existing measurements of the spectrum at EeV energies.
Note that at $\sim 3$~EeV the normalisation of the HiRes and AGASA  spectra
are above the one reported by Auger. In particular, the HiRes normalisation
would correspond to adopting $\kappa=1.2$  while the AGASA normalisation would
correspond to a value for $\kappa$ of about two.

Consider a
Gaussian filter matching the angular resolution characterized by $\sigma$
\begin{equation}
W(\beta)\equiv \exp\left(-\frac{\beta^2}{2\sigma^2}\right),
\end{equation}
where $\beta$ is the angle from the direction of Sagittarius $A^*$. 
Then the expected number of events in the specified energy range is 
\begin{equation}
n_{exp}=2\pi\int_0^{\pi}  {\rm d}\beta \ \sin\beta\;W(\beta)\int_{E_{min}}^{E_{max}}
{\rm d}E\ A(E) \Phi_{CR}(E),
\end{equation}
where $A(E)$ is the energy dependent exposure of the experiment. 
Similarly,
the number of events expected to be observed from the point-like source
will be
\begin{equation}
n_s=\int_0^{\pi} \frac{{\rm d}\beta\;\sin\beta}{\sigma^2}
W(\beta)^2\int_{E_{min}}^{E_{max}} 
{\rm d}E\ A(E) \Phi_s(E),
\end{equation}
where we take into account that, due to the finite angular resolution of the
experiment,  the 
arrival directions of the observed source events are
expected to be distributed according to
\begin{equation}
\frac{{\rm d}\Phi_s}{{\rm
    d}\Omega}(\beta,E)=\frac{\exp\left(-\beta^2/2\sigma^2\right)}{2\pi\sigma^2}
    \Phi_s(E).
\end{equation}

Using the assumptions noted above, we then get an expression
for the source flux 
integrated over the
energy range considered,  
\begin{equation}
\Phi_s\equiv \int_{E_{min}}^{E_{max}} 
{\rm d}E\ \Phi_s(E)
\end{equation}
 with a 95\% CL  upper bound of 
\begin{equation}
\Phi_s^{95}=\frac{n_s^{95}}{n_{exp}}4\pi\sigma^2
\int_{E_{min}}^{E_{max}}{\rm d}E
\ \Phi_{CR}(E)=\kappa\; 0.13 \ {\rm km^{-2}\ yr^{-1}}.
\label{sdbound}
\end{equation}

Note that the bound on the source flux just scales with the parameter 
$\kappa$,
because what is constrained is the ratio between the source and background
fluxes. 

Let us now discuss how the bound would change if the bulk of the 
CRs were heavy nuclei in this
energy range. Following the discussion in the previous Section, 
 we conclude that the upper limit to the flux from the putative source
will have to be scaled by a factor $\sim
1.3$ under the assumption that the CRs are iron nuclei and that the source 
is a source of neutrons.  We thus see
 that the bound on the neutron flux could be up to 
$\sim 30$\% higher if the CR composition at EeV energies were
 heavy.

Due to the steeply falling CR spectrum, the bound in eq.~(\ref{sdbound}) also
holds for $E_{max}\to \infty$, i.e. in the inclusive range
$E>10^{17.9}$~eV. Setting instead $E_{min}=1$~EeV, the corresponding
bound is $\Phi_s^{95}=\kappa\; 0.06 \ {\rm km^{-2}\ yr^{-1}}$.

We point out that some of the theoretical predictions for neutron
fluxes (those associated with the AGASA claim, but not those associated with
 the
TeV results) are based on
 the AGASA normalization for the CR flux, which is about a
factor of 3 larger than the Auger flux normalization. The earlier predictions 
 must thus be reduced by 
this factor to be compared with the
flux bounds obtained here.
 The predictions of refs.~\cite{bo03}, \cite{ah05}
 and  \cite{cr05}, which exceed the
upper-bound obtained by more than one order of magnitude, 
 are already excluded, and that of \cite{gr05} is 
at the level of the present Auger sensitivity.

\subsubsection{The hybrid results}

We have 
also studied the GC region as observed with hybrid events, detected by both
the FD and SD. These events have a better angular resolution \cite{angres}
($ 0.7^\circ$ at 68\% C.L. in the energy range studied). 

Considering the events with  $10^{17.9}\ {\rm eV}<E<10^{18.5}$~eV,
 no significant excess is seen in the GC direction. 
For instance, in an optimal top-hat window of
$1.59\sigma\simeq 0.75^\circ$ radius, 
 0.3 events are expected (as estimated using a shuffling method)  while 
no single event direction falls within that circle.
This leads to a source flux upper-bound at 95\% CL of
\begin{equation}
\Phi_s^{95}=\kappa\; 0.24 \ {\rm km^{-2}\ yr^{-1}},
\end{equation}
which is about a factor of two weaker than the SD flux bound.
Note that
the energy assignments of the FD apply regardless of the assumed CR
composition (except for a small correction to account for the missing energy),
be they protons or heavy nuclei. However,
the acceptance has  a dependence on composition because different primaries
develop at different depths in the atmosphere. Since a quality
requirement for hybrid events is to have the maximum of the shower
development inside the field of view of the telescopes, this affects
the sensitivity to different primaries.
The bound obtained is indeed a conservative one if the bulk of the CRs are
heavy nuclei.

\subsubsection{Relation to a point-like photon source}

In \cite{hess} the H.E.S.S. collaboration has reported a remarkably flat
 spectrum of gamma rays above 165~GeV (and up to 10~TeV) 
from the direction of Sagittarius
 A$^*$. A naive extrapolation of this spectrum would lead to 
a flux of gamma rays
 above 1~EeV of $~0.04  \ {\rm km^{-2}\ yr^{-1}}$. 
Note however that the
 bound obtained by us for a neutron source (which is comparable to this
 extrapolation) does not apply straightforwardly for photon primaries, 
since the acceptance (and energy assignments) are modified.

The spectrum of photons reported from the GC ridge \cite{hess2} is 
also remarkably flat so that this region too merits future study.
The Galactic Centre may house sources of very high-energy cosmic rays
 detectable through gamma radiation. 
It is clear then that further exposure with
 the Auger Observatory of this region and a  dedicated analysis 
will be of interest.  Also an exploration down to the FD threshold will be
important for the search of photon sources.

\section{Conclusions}

Using the first 2.3 years of Auger data we have searched for localized
anisotropies near
the direction of the Galactic Centre, which is well within the field of view
of the Observatory. 
With statistics much greater than those of
previous experiments, we have looked for a point-like source in the
direction of Sagittarius $A$, without finding a significant excess.
 This
 excludes several scenarios of neutron sources in the GC suggested recently.
 Our searches on larger angular windows in the neighborhood of the GC do not
 show abnormally over-dense regions.
In
particular, they do not support the large excesses reported in AGASA
data (of 22\% on $20^\circ$ scales) and SUGAR data 
(of 85\% on $5.5^\circ$ scales).

\section*{Acknowledgments}
We are very grateful to the following agencies and organizations for
financial support:
Gobierno de Mendoza, Comisi\'on Nacional de Energ\'\i a At\'omica y 
Municipalidad de Malarg\"ue, Argentina;
the Australian Research Council;
Fundacao de Amparo a Pesquisa do Estado de Sao Paulo,
Conselho Nacional de Desenvolvimento Cientifico e Tecnologico and
Fundacao de Amparo a Pesquisa do Estado de Rio de Janeiro, Brasil;
National Science Foundation of China;
Ministry of Education of the Czech Republic (projects LA134 and
LN00A006);
Centre National de la Recherche Scientifique, Institut National de
Physique Nucl\'eaire et Physique des Particules (IN2P3/CNRS),
Institut National des Sciences de l'Univers (INSU/CNRS) et Conseil
R\'egional Ile de France, France;
German Ministry for Education and Research and Forschungszentrum
Karls\-ruhe, Germany;
Istituto Nazionale di Fisica Nucleare, Italy;
Consejo Nacional de Ciencia y Tecnologia, Mexico;
the Polish State Committee for Scientific Research (grant numbers
1P03D~01430, 2P03B~11024 and 2PO3D~01124), Poland;
Slovenian Research Agency; 
 Ministerio de Educaci\'on y Ciencia
(FPA2003-08733-C02, 2004-01198),
Xunta de Galicia (2003 PXIC20612PN, 2005 PXIC20604PN) and Feder Funds,
Spain;
Particle Physics and Astronomy Research Council, UK;
the US Department of Energy, the US National Science Foundation, USA,
and UNESCO.

\end{document}